# Quantifying the Blue Shift in the Light Absorption of Small Gold Nanoparticles

R. Tsekov, P. Georgiev, S. Simeonova and K. Balashev
Department of Physical Chemistry, University of Sofia, 1164 Sofia, Bulgaria

The dependence of the surface plasmons resonance (SPR) frequency on the size of gold nanoparticles (GNPs) is experimentally studied. The measured data for the SPR frequency by UV-Vis spectroscopy and GNPs diameter by Dynamic Light Scattering (DLS), Transmission Electron Microscopy (TEM) and Atomic Force Microscopy (AFM) are collected in the course of classical citrate GNPs synthesis. The relationship between the GNPs size and the blue shift of the light absorption is presented. They are fitted by an equation with a single free parameter, the dielectric permittivity of the surrounding media. Thus, the refractive index of the surrounding media is determined, which characterizes the GNPs surface shell.

It is well known that the optical properties of metal nanoparticles, such as Au, Ag, Cu, etc., with sizes less than 20 nm are determined by the collective oscillation of valence electrons [1]. These electrons interact with the electric field of the incident radiation, which induces a dipole in the nanoparticle. The optical properties of metal nanoparticles' dispersions can be quantified by studying the intensity of the absorption maximum due to surface plasmon resonance (SPR), measured from the UV-Vis spectrum. The application of the UV-Vis spectroscopy for investigation of gold nanoparticles growth is possible, because the absorption intensity of the SPR maxima depends on the nanoparticle size [1-4]. Drude and Lorentz have derived the fundamental laws for the optical properties of bulk metals [5]. Their models provide the frequency dependence of the dielectric function of a metal. The Drude model is employed for description of the dielectric function of metals with free valence electrons such as gold, silver and copper [6]. Johnson and Christy [7] applied the Drude model for description of the ellipsometric measurements of the dielectric function of gold, silver and copper films. Recently, Losurdo et al. [8] measured the ellipsometric spectra of the GNPs films and the sizes of the GNPs were characterized by AFM. The experimental results showed dependence of the dielectric function from the size of GNPs, organized in the films. The obtained experimental dependence were parametrized by a combination of three Lorentzian oscillators. Jenkins et al [9] studied blue shifted narrow localized surface plasmon resonance (LSPR) from dipole coupling in GNPs random arrays. The LSPR width is narrower than that of the single gold nanoparticles. The blue-shifted LSPR is due to the long-range dipole coupling in the gold nanoparticle random arrays indicated from simulations using the T-matrix method.

The absorption of small gold nanoparticles in solution and extinction cross section for single spherical particle were calculated via the Mie equation [3], where for the dielectric function was used a semi-empirical equation. Accordingly, the theoretical predictions for small size GNPs (up to 20 nm) gave the increase of the intensity of the absorption maxima with respect to the increase of the GNPs size. Although this theoretical approach was not capable of explaining the blue shift in the UV-Vis spectra, which is experimentally observed in the course of the GNPs synthesis correspondingly to the increasing GNPs size [3, 4]. By taking into account the blue shift we propose a simple relation, where the surface plasmon resonance frequency is a function of the GNPs' size. In order to verify the derived equation for the plasmon frequency and for the nanoparticles sizes, kinetic data obtained in the course of their synthesis are used. For this purpose, the DLS, TEM and AFM size distributions are analyzed and the SPR frequencies (from UV-Vis spectra) and corresponding GNPs sizes, could be determined by the proposed simple equations.

**Experimental**

*Chemicals and reagents:* Analytical grade Tetrachloroauric acid ($HAuCl_4 \cdot 3H_2O$) was purchased from Panreac Química (Spain), Trisodium citrate ($Na_3C_3H_5O(COO)_3 \cdot 2H_2O$) and Sodium chloride (NaCl) were obtained from Sigma-Aldrich (Germany). All solutions were prepared in deionized water.

*Synthesis of spherical gold nanoparticles:* The GNPs synthesis procedure, as previously described [3, 4], was based on citrate method [12], one of the best methods for synthesis of monodisperse spherical GNPs is as follows: 10 ml solution of tetrachloroauric acid, containing 5 mg gold, was added to 85 ml deionized water. The mixture was stirred at 350 rpm and heated, and 5 ml 1% solution of trisodium citrate was added as reducing and stabilizing agent. The excess of trisodium citrate concentration is for stabilizing action – formation of citrate shell on nanoparticles surface. The synthesis were performed at two reaction temperatures, 70 °C and 90 °C.

*Experimental methods:* The UV-Vis absorption spectra were measured by spectrophotometers Thermo Scientific Evolution 300 UV-Vis Spectrophotometer (Fisher Scientific, USA) and Jenway 6400 (Krackeler Scientific Inc., USA). Additional blue shifted UV-Via spectra are published in previous works [3, 4]. The size distribution of gold nanoparticles was determined by Dynamic Light Scattering (DLS) apparatus Zetasizer Nano ZS (Malvern Instrument Ltd., Malvern, UK). The setup was equipped with 532 nm HeNe gas laser and detector optics and an ITT FW 130 photomultiplier and ALV-PM-PD amplifier-discriminator. The morphology and the size of GNPs were determined by JEM-2100 LaB6 Transmission Electron Microscope (JEOL Ltd., Japan), and TEM Philips CM-10, used at 100kV. The sample preparation for AFM imaging is described elsewhere [3, 4]. AFM imaging was performed on NanoScope V system (Bruker Inc.) operating in tapping mode in air at room temperature with silicon cantilevers (Tap300Al-G, Budget Sensors, Innovative solutions Ltd., Bulgaria). The NanoScope software was used for section analysis and particle size determination [3, 4]. All the methods were used in the course of the GNPs synthesis as the samples were taken at different reaction time intervals [3, 4]. The GNPs size distributions from TEM micrographs and AFM images were obtained from analysis of about 200 nanoparticles of each of the samples taken during the synthesis.

*Experimental observations:* The 2D AFM images, presented in Fig. 1A,D and corresponding 3D AFM images, were obtained at 30[th] and 70[th] min from the GNPs synthesis, performed at 70 °C, after depositing some amount of the gold suspension on mica support. The particle diameter were determined by section analysis, performed across certain imaged gold nanoparticles on Fig. 1B,E. Representative images of gold nanoparticles, taken with High Resolution (HR) TEM, are shown in Fig. 1G,H. The figure represents gold nanoparticles synthesized by us at temperature 90 ºC at the final stage of synthesis. The nanoparticles are mostly spherical. Aggregates of nanoparticles could also observed to form when the limited protection ability of sodium citrate, which serves not only as a reducing agent, but also as a stabilizing one.

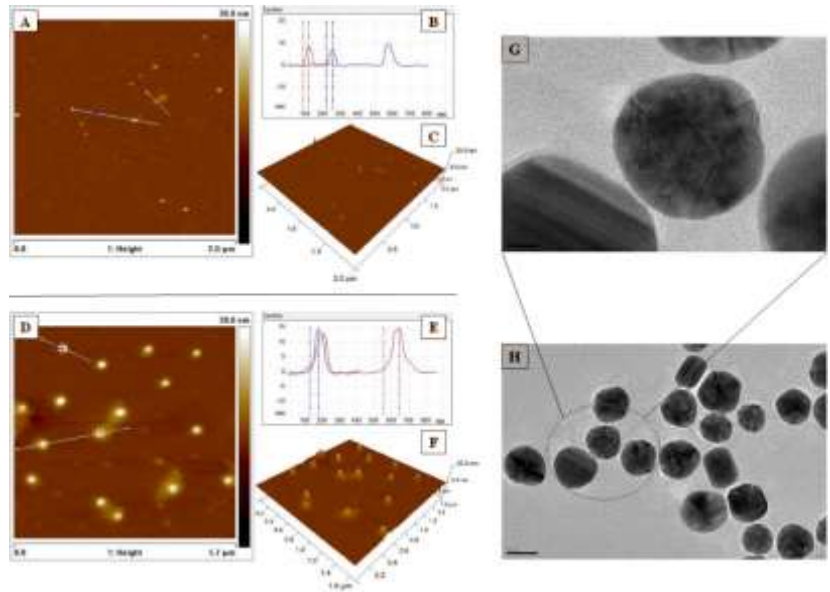

**Fig. 1.** (A, D) 2D AFM images with (B,E) section analysis across the imaged GNPs are z=10 nm and z=15 nm, (C,F) corresponding 3D AFM images. The z size of the AFM instrumental error is *d*=±0.2 nm. (G, H) HR-TEM micrographs of GNPs at the final stage of growth. The scale bars are 5 nm and 20 nm. The resolution of the TEM is *d*=±0.19 nm.

It is well-known that in the process of GNPs' growth the number of particles stays nearly constant, equal to the number of initial nuclei [3, 9, 13-16]. In agreement with this postulate it is expected that the extinction cross section of the single GNP. Since the absorbance is proportional to the extinction cross section, this explains the increase of the absorbance with the increase of the nanoparticles size. The blue shift of the absorption maxima and the change of the extinction cross section in the course of GNPs growth are presented at Fig. 2A while at Fig 2B are presented the histograms of the GNPs determined from AFM (Fig. 2B). Within the instrumental error of UV-Vis and AFM data it is possible to quantify the blue shift of the plasmon maxima of the GNPs with diameters up to 20 nm. The comparison of experimental absorption spectra with extinction cross section gives proportionality constant [3]. This constant $A/C_{ext} = 3.1 \times 10^{15}$ is very close to one calculated for an idealized case of monodisperse solution of GNPs with spherical shape $A/C_{ext} = 2.7 \times 10^{15}$ [3]. This result is in agreement with well-known formula for absorbance of colloidal suspension $A/C_{ext} = N_{NP}L/2.3$, where $N_{NP}$ is the number of nanoparticles per unit volume and *L*=1 cm the light path length [2, 3]. The observed blue shift is considered to be a function only of GNPs size and not of the number of gold nanoparticles, because experimental measured spectra for different concentrations has a constant wavelength (data not shown). The absorption spectra can be described by the equation Eq. (1) giving the relation between the extinction cross section and fitting variables - GNPs' size and the wavelength. Here were propose a simple relation between the blue shifted plasmon absorption maxima (wavelength) and the size of GNPs which have sizes smaller than 20 nm. Fig. 2B represents the AFM size distributions of GNPs corresponding to the absorbance spectra of the same GNPs samples taken in the course of their synthesis. The curves described the histogram was made by plotting the number of particles with a specified average size as a function of their diameter *d*. The mean diameter, was the value of the maximum of size distribution. It is clear from the distributions in Fig. 2B that the size distribution does not change appreciably for different samples. The same conclusions could be made from DLS size distributions data (data not shown). Having measured of absorbance spectra, we can relate them to the GNPs' size with a simple equation.

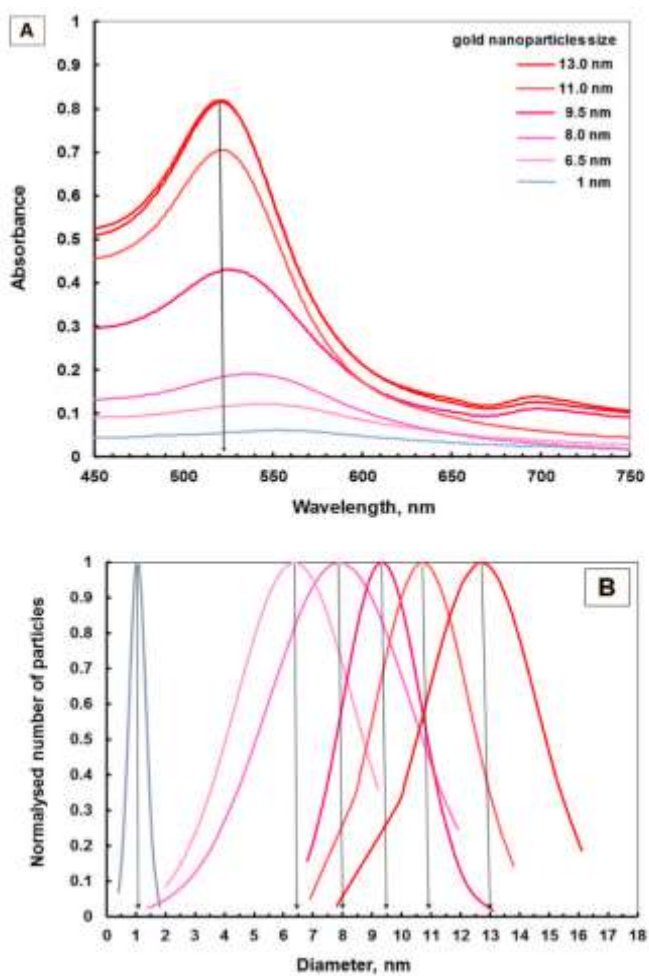

**Fig. 2.** (A) Blue shift of the spectra during the citrate synthesis performed at 70 °C and (B) The histograms of the corresponding GNPs size distributions derived from AFM data. The instrumental error of UV-Vis spectrophotometer wavelength is $\lambda=\pm 1$ nm, and the instrumental error of AFM for each GNP size determination is $d=\pm 0.2$ nm.

**Theoretical Results and Discussions**

It is well known that incident light excites surface plasmons in gold nanoparticles. According to the Mie equation, the extinction cross-section of a single particle is given by [2-4]

$$C_{ext} = \frac{3\pi^2 d^3 \varepsilon_m^{3/2}}{\lambda} \frac{\varepsilon_{Im}}{(\varepsilon_{Re} + 2\varepsilon_m)^2 + \varepsilon_{Im}^2} \qquad (1)$$

where $d$ is the nanoparticle diameter, $\lambda$ is the light wavelength, $\varepsilon_m$ is the dielectric constant of the medium, $\varepsilon_{Re}$ and $\varepsilon_{Im}$ are the real and imaginary parts of the dielectric function of metal nanoparticles, respectively. A typical value of the dielectric constant of water is $\varepsilon_m = 1.775$. The relevant frequency dependence of the nanoparticle dielectric permittivity is described by the adapted Drude formula [17]

$$\varepsilon(\omega) = 1 - \frac{\omega_{sp}^2}{\omega^2 - i\omega\gamma} \qquad (2)$$

where $\omega_{sp}$ is the surface plasmon frequency and $\gamma$ is a damping coefficient. One can easily obtain from Eq. (2) that $\varepsilon_{Re} = 1 - \omega_{sp}^2/(\omega^2 + \gamma^2)$. Substituting it in the resonant condition $\varepsilon_{Re} + 2\varepsilon_m = 0$, following from Eq. (1), yields the frequency $\omega_{max} \equiv 2\pi c/\lambda_{max}$ of the spectral peak

$$\omega_{max}^2 = \frac{\omega_{sp}^2}{2\varepsilon_m + 1} - \gamma^2 \qquad (3)$$

where $c = 300$ Mm/s is the speed of light in vacuum. The dielectric constant $\varepsilon_m$ relates also the surface plasmon frequency $\omega_{sp}^2 = \omega_p^2/(\varepsilon_m + 1)$ to the bulk plasmon frequency $\omega_p$, which is equal to $\omega_p = 13.7$ PHz for gold.

The mean free path of electrons in gold is $l = 38$ nm [6]. This means that in smaller gold nanoparticles the dissipation of energy occurs mainly by collisions of electrons on the particle surface, similar to the Knudsen diffusion. Hence, the corresponding damping coefficient should be proportional to the Fermi velocity divided by the particle diameter $d$ [17]. Since the plasmon wave vector is inversely proportional to $d$, the size dependence of the damping coefficient can be modelled well by $\gamma = \gamma_\infty + A v_F/d$, where $v_F = 1.39$ Mm/s is the Fermi velocity for gold. Because the dimensionless parameter $A$ for bulk is $2\pi$, the damping coefficient of the bulk gold can be estimated as $\gamma_\infty = 2\pi v_F/l = 0.23$ PHz. Analyzing further Eq. (1) unveils that the damping coefficient is approximately equal to spectrum width at half maximum, i.e. $\gamma \approx \Delta_{1/2}$. The symmetrized width at half maximum of the experimental spectra is plotted in Fig. 3 as a function of the reciprocal GNPs diameter. In Fig. 3 the experimental data are plotted in accordance to Eq. (3) as open circles refer to synthesis at temperatures 90 °C from TEM (Fig. 4A), DLS (Fig. 4B), and AFM (Fig. 4C) also at 70 °C (black circles) [3, 4].

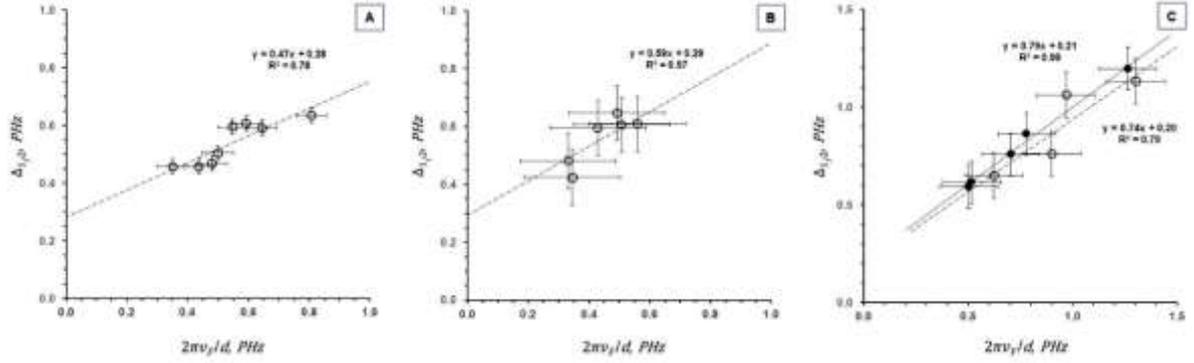

**Fig. 3.** The symmetrized width at half maximum $\Delta_{1/2}$ PHz of the experimental spectra as a function of the reciprocal particle diameter $2\pi v_F/d$ PHz. The data for $\Delta_{1/2}$ are obtained from UV-Vis spectra during the citrate synthesis performed 90 °C (open circles) and 70 °C (black circles). The data for GNPs diameter are obtained from size distribution maxima: (A) TEM; (B) DLS; (C) AFM. The error bars represent the standard deviation.

The comparison of slopes and intercepts from linear fits in Fig. 3 are presented at Table 1. All of the applied experimental methods: TEM, DLS and AFM gives similar values from theoretical fits.

**Table 1**

| Linear fit | Temperature | TEM | DLS | AFM |
| --- | --- | --- | --- | --- |
| Slope | 90 °C | 0.47 | 0.59 | 0.74 |
| Intercept | 90 °C | 0.28 | 0.29 | 0.20 |
| Slope | 70 °C | × | × | 0.79 |
| intercept | 70 °C | × | × | 0.21 |

As is seen, the theoretical prediction describes well the experimental data. Using the relation above, one can replace the damping coefficient $\gamma$ with the experimentally measured $\Delta_{1/2}$ in Eq. (3) to obtain after some rearrangements the following equation:

$$\frac{\omega_{max}^2 + \Delta_{1/2}^2}{\omega_p^2} = \frac{1}{(2\varepsilon_m + 1)(\varepsilon_m + 1)} \quad (4)$$

It shows a fundamental relationship between the resonant frequency $\omega_{max}$ of the spectral peak and the width $\Delta_{1/2}$ at half maximum: the sum $\omega_{max}^2 + \Delta_{1/2}^2$ is independent on the particle diameter $d$. Since $\Delta_{1/2}$ decreases with particle size (see Fig. 3), it follows that $\omega_{max}$ experiences a blue shift with increase of $d$.

In Fig. 4 the experimental data are plotted in accordance to Eq. (4) as open circles refer to synthesizes at temperatures 90 °C from TEM (Fig. 4A). DLS (Fig. 4B), and AFM (Fig. 4C) also at 70 °C (black circles) [3, 4]. As is seen, Eq. (4) holds firm and one can calculate the value $\varepsilon_m = 1.897$ of the medium dielectric constant from $1/(2\varepsilon_m + 1)(\varepsilon_m + 1) = 0.072$. Using the relation $n = \varepsilon_m^{1/2}$ this value can be easily transformed in the relevant refractive index $n = 1.377$. The measured refractive index of the initial solution is $n = 1.335$, which is close to $n = 1.333$ of pure water, since the solutions are relatively dilute. Since the value of the refractive index of pure sodium citrate is $n = 1.394$, the obtained higher value of the refractive index can be attributed to adsorption of citrate anions on the GNPs (c.a. 70%), leading to stabilization of the suspension.

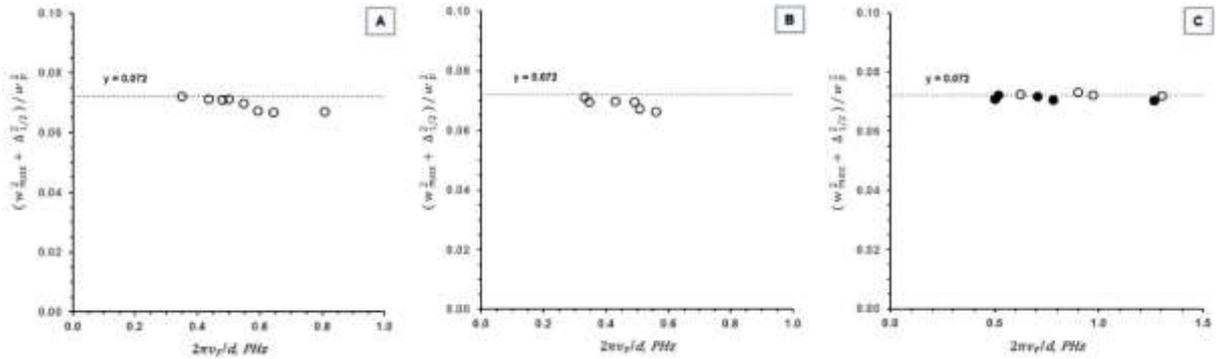

**Fig. 4.** The experimentally determined $(\omega^2_{max}+\Delta^2_{1/2})/\omega^2_p$ plotted as a function of the reciprocal particle diameter $2\pi v_F/d$ PHz. The meaning of the symbols is the same as in Fig. 3. The data for $\Delta_{1/2}$ are obtained from UV-Vis spectra during the citrate synthesis performed at temperature 90 °C (open circles) and 70 °C (black circles). The data for GNPs diameter are obtained from size distribution maxima: (A) TEM; (B) DLS; (C) AFM.

### Conclusions

In additional discussion for size and shape there are two plasmon peaks in the absorbance spectra gold nanorods, formed by growth on spherical citrate stabilized GNPs seeds. The first one corresponds to the so-called transversal plasmon, localized with respect to the short axis of ellipsoid. It stays constant (525 nm) to short axis. The second peak corresponds to the longitudinal plasmon, localized with respect to the long axis of ellipsoid. With decreasing the amount of spherical citrate stabilized GNPs seeds, the length of ellipsoid is increasing. In this case longitudinal plasmon is red shifted from 750 to 980 nm with increasing the long axis [18]. The extinction cross section equation for generation of surface plasmons in an ellipsoid [1], in difference to sphere, include depolarization factors for the three axes of the ellipsoid. In conclusion the blue shift effect was observed is not to anisotrotropic or tailored GNPs size and shape, or due to aggregation of GNPs.

The fundamental Eq. (3), relating the SPR frequency to the GNPs size via the damping parameter, describes the blue shift of the light absorption. The analytical dependence $\omega_{max}(d)$ has only one fitting parameter, the dielectric permittivity of the surrounding GNPs media. Thus Eq. (4) was applied for fitting the experimental data extracted from UV-Vis spectroscopy and AFM images in the course of GNPs synthesis. From the experimental data and the theoretical fit, the refractive index values of the surrounding GNPs media was determined, which provides valuable information for the adsorption shell (citrate) of the gold nanoparticles.